\documentclass[twocolumn]{IEEEtran}

\usepackage{amsfonts}
\usepackage{amssymb}
\usepackage{graphicx}
\begin{document}
			   
\title{Atomistic investigation of low-field mobility in graphene nanoribbons}
\author{Alessandro Betti, Gianluca Fiori and Giuseppe Iannaccone\\
Dipartimento di Ingegneria dell'Informazione: Elettronica, Informatica, Telecomunicazioni, \\
Universit\`a di Pisa, via Caruso 16, 56100 Pisa, Italy\\
email: {\{alessandro.betti, g.iannaccone\}@iet.unipi.it, gfiori@mercurio.iet.unipi.it}, Tel. +39 050 2217639}

\maketitle

\bibliographystyle{unsrt}



\begin{abstract}
We have investigated the main scattering mechanisms affecting mobility
in graphene nanoribbons using detailed atomistic simulations.
We have considered carrier scattering due to acoustic and optical phonons, 
edge roughness, single defects, and ionized impurities, and 
we have defined a methodology based on 
simulations of statistically meaningful ensembles of nanoribbon segments. 
Edge disorder heavily affects mobility at 
room temperature in narrower nanoribbons, whereas charged impurities and
phonons are hardly the limiting factors. Results are favorably compared
to the few experiments available in the literature.

\end{abstract}


\IEEEpeerreviewmaketitle
\IEEEoverridecommandlockouts

\section{Introduction} 

Two-dimensional (2D) graphene sheets have 
demonstrated really attractive electrical properties like high 
carrier mobility~\cite{Chen,Li} and large coherence 
length~\cite{Appenzeller}. 
However, experimental data of mobility available in the literature show 
huge dispersion, ranging from 10$^2$ to 10$^4$~cm$^2$/Vs at room temperature, 
signaling that the fabrication process is still poorly optimized and not fully
repeatable. To guide process optimization, an exhaustive interpretation of 
physical mechanisms limiting mobility would be extremely useful.
For Graphene NanoRibbons (GNRs) a comprehensive experimental
characterization of mobility is still lacking, mainly due to the
difficulty in patterning in a repeatable way very narrow ribbons. Few
recent interesting experiments are reported in~\cite{Wang} and~\cite{Yang}.
GNRs may also suffer significant degradation of mobility due to 
additional scattering mechanims, such as edge roughness.

The single most important aspect that makes graphene interesting for 
nanoscale electronics is its very high mobility. It is therefore of paramount 
importance to 
understand if also nano structured graphene can preserve the high mobility 
(often) measured in graphene sheets, much larger
than that of conventional semiconductors. In addition, one would need to 
understand the effect on mobility of different options for graphene 
functionalization, which could be required to open a semiconducting gap 
in graphene.

In the current situation, theoretical investigations~\cite{Fang,Bresciani} and numerical 
simulations~\cite{Areshkin,Querlioz,BettiIEDM09} 
can represent a useful tool to assess the relative impact of 
different sources of non-idealities on mobility and consequently on 
device performance, to provide guidelines for the fabrication process
and a realistic evaluation of the perspectives of graphene in nanoelectronics.

An analytical method and a Monte Carlo approach have for example been 
adopted in order to study line-edge roughness 
(LER) and phonon scattering-limited mobility in Ref.~\cite{Fang} 
and Ref.~\cite{Bresciani}, respectively. 
However, due to the reduced width of the considered devices, effects 
at the atomistic scale are relevant, therefore accurate simulation 
approaches like semi-empirical tight-binding are needed. 

In this work we present atomistic simulations of GNR-FETs, considering 
GNR widths ranging from 1 to 10~nm, and 
including scattering due to LER, single defects, ionized 
impurities, acoustic and optical phonons. 
A direct comparison with recently fabricated 
devices~\cite{Wang} will also be performed. 
Statistical simulations performed on a large ensemble of nanoribbons with
different occurrences of the spatial distribution of non-idealities show that phonons, LER and defects 
scattering can likely explain the few available experimental 
data~\cite{Wang}, where mobility is down to the level of mundane semiconductors (order of $10^2$-$10^3$~cm$^2/$Vs). 

\section{Methodology} 

\begin{figure} [tbp] 
\begin{center} 
\includegraphics[width=7.0cm]{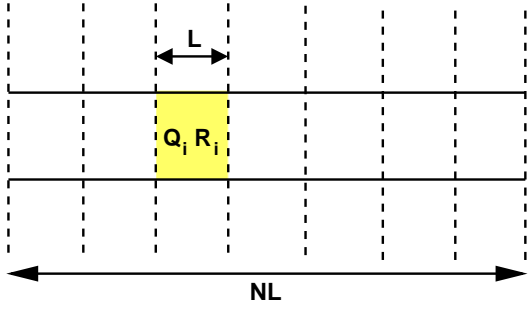}
\end{center}
\vspace{-0.2cm}
\caption{GNR-FET channel with length $NL$ and the simulated GNR segment with 
length $L$.} 
\label{fig:segment}
\end{figure}

A long GNR-FET channel, where mobility is properly defined, is given by a 
series of $N$ GNR segments of length $L$ like those we have 
considered in the simulation (Fig.~\ref{fig:segment}).
For the $i$-th GNR segment, the resistance $R_i=V_{DS}/I_i$ 
is the sum of two contributions, the channel resistance $R_{ch,i}$ 
and the contact (ballistic) resistance $R_B=V_{DS}/I_B$ 
($R_i=R_{ch,i}+R_B$), 
where $V_{DS}$ is the drain-to-source voltage, whereas $I_i$ and $I_B$ are 
the total current and the ballistic current in the $i$-th segment, 
respectively. 
Assuming phase coherence is lost at the interface between segments, 
the resistance $R_{tot}$ of the long channel GNR is therefore 
the sum of $N$ channel resistances and one contact resistance, i.e.:

\begin{eqnarray} \label{eqn:totalresistance1}
R_{tot}=\left(\sum_{i=1}^N R_{ch,i}\right) + R_B= N \langle R \rangle -\left(N-1\right) R_B \, ,
\end{eqnarray}
where $\langle R \rangle= (1/N) \,\sum_{i=1}^N R_i$ is the mean resistance 
evaluated on the ensemble of nanoribbon segments. 
Therefore, the mobility of a long channel would read:
\begin{eqnarray} \label{eqn:mobility}
\mu_n=\frac{L_{tot}^2 \, G_{tot}}{Q_{tot}}=\frac{L_{tot}^2}{Q_{tot}} \frac{1}{N \langle R \rangle - \left(N-1 \right) R_B} \, ,
\end{eqnarray}
where the index $n$ denotes each type of scattering mechanism 
limiting mobility (defects, edge-roughness or impurities), $L_{tot} = NL$ is 
the total GNR length, $Q_{tot}=\sum_{i=1}^N Q_i = N \, \langle Q \rangle$ is the total charge 
in the channel and $\langle Q \rangle $ is the mean mobile charge in a segment. 

For large values of $N$, one can discard 1 with respect to $N$ 
in Eq.~(\ref{eqn:mobility}) so that we obtain the formula we use in the 
paper~\cite{IannaPala}:
\begin{equation} \label{eqn:mobility1}
\mu_n=\frac{L^2}{(\langle R \rangle - R_B) \langle Q \rangle}\, ,
\end{equation}
The root mean square error of mobility $\sigma_\mu$ 
has been computed by means of a 
Taylor expansion up to the first order of Eq.~(\ref{eqn:mobility1}) with 
respect to statistical fluctuations of the resistance $R= R_{ch}+R_B$:
\begin{eqnarray} \label{eqn:error}
\Delta \mu= \left|\frac{\partial \mu}{\partial R} \right| \Delta R= \frac{L^2}{\langle Q \rangle} \frac{\Delta R}{\left(\langle R \rangle - R_B \right)^2}= \mu \frac{\Delta R}{\langle R\rangle - R_B} \, ,
\end{eqnarray}
and therefore
\begin{equation} \label{eqn:error1}
\sigma_\mu^2 = \left( \frac{\mu}{\langle R\rangle - R_B} \right)^2 \sigma_R^2
\end{equation}
where $\Delta R= \sqrt{\sigma_R^2/N}$ and 
\begin{eqnarray} \label{eqn:error2}
\sigma_R^2= 1/\left(N-1 \right) \sum_{i=1}^N \left(R_i - \langle R \rangle \right)^2 
\end{eqnarray}
is the variance of $R$.

Statistical simulations of resistance 
on a large ensemble of nanoribbon segments with different actual 
distribution of non-idealities have been performed.  
In particular, the mobility $\mu_n$ has been computed in 
the linear transport 
regime, for large gate voltages ($V_{GS}$) and small drain-to-source 
bias $V_{DS}=$~10 mV. 
Mobility has been extracted by means 
of Eq.~(\ref{eqn:mobility1}) considering an ensemble of 
$N=600$ nanoribbon segments with different disorder realizations 
for 1.12 nm-wide GNRs. Due to the computational cost, at least 40 
nanoribbons segments have been instead simulated for 10.10 nm-wide GNRs. 

Statistical simulations of random actual distributions of defects, LER 
and ionized impurities have been computed through the self-consistent solution 
of 3D Poisson and Schr\"odinger equations within the NEGF formalism, with a
$p_z$ tight-binding Hamiltonian~\cite{BettiIEDM09}, extensively 
exploiting our open-source simulator NanoTCAD ViDES~\cite{ViDES}. 
In particular, we have imposed at both ends of the segments 
null Neumann boundary conditions on the 
potential, and open boundary conditions for the transport equation. 

In order to compute the LER-limited mobility $\mu_{LER}$, 
statistical simulations have been performed 
considering a given fraction $H$ of single vacancy defects at the edges.
$H$ is defined as the probability for each carbon atom at the edges
to be vacant. In practice, each sample of nanoribbon 
with edge disorder is randomly generated 
assuming that each carbon site at the edges has a probability $H$ to be replaced by a vacancy.
Null hopping parameter has been imposed in correspondence of a defect 
at the edge. 

Defects have been modeled using the on-site energy and 
the hopping parameter extracted from DFT 
calculations~\cite{Deretzis}. 
In particular, for a fixed defect concentration $n_d$, each 
sample of defected nanoribbon
with defects is randomly generated assuming that each carbon atom
has a probability $n_d$ to be replaced by a vacancy.

As previously assumed in ab-initio 
calculations~\cite{Rytkonen}, we have 
considered a surface impurity distribution of positive charges equal to 
+0.4~$q$ placed at a distance of 0.2~nm from the GNR surface, where $q$ is the 
elementary charge. 
Again, if  $n_{IMP}$ is the impurity fraction, a sample 
with surface impurities is randomly generated by assuming 
that each carbon atom has a probability $n_{IMP}$ 
to be at 0.2~nm from an impurity in the dielectric layer.

\begin{figure} [tbp] 
\begin{center} 
\includegraphics[width=7.0cm]{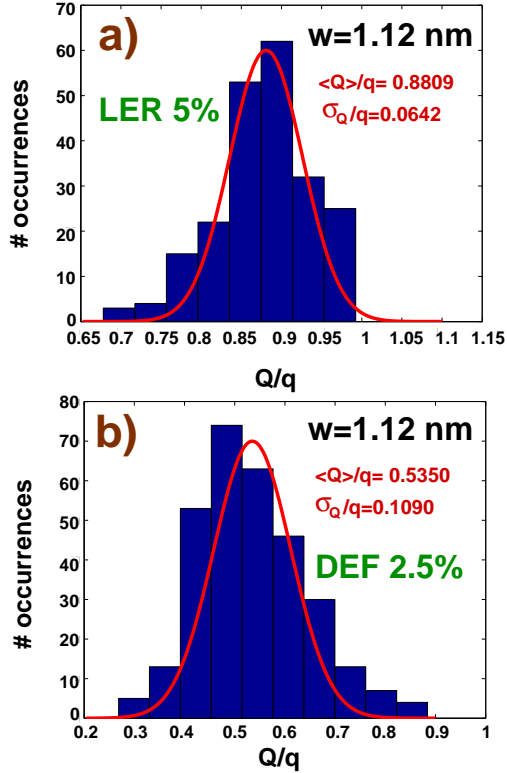}
\end{center}
\vspace{-0.6cm}
\caption{The distributions of charge $Q$ in each GNR segment 
($W=$~1.12~nm) obtained 
from statistical simulations of random distributions of (a) LER 
($H=$~5\%) and (b) defects ($n_d=$~2.5\%).} 
\label{fig:istogramma}
\end{figure}

In Figs.~\ref{fig:istogramma}a-b, we show the distributions of $Q$ 
when considering line-edge roughness ($H=$~5\%) and 
defects ($n_d=$~2.5\%) for $W=$~1.12~nm. In each picture we show the 
mean value $\langle Q \rangle$ and the standard deviation $\sigma_Q$ of 
the random variable $Q$. For comparison, the corresponding normal distribution is shown.
 
Phonon-limited mobility $\mu_{ph}$ 
(both acoustic and optical) has been  computed by means of a 
semi-analytical model as in~\cite{Fang}, but extending the Kubo-Greenwood 
formalism beyond the effective mass approximation and 
accounting for energy relaxation at GNR edges~\cite{Michetti}. 
Starting from the Boltzmann transport equation, the phonon limited 
mobility for a 1D conductor can be espressed as~\cite{Fischetti}~:
\begin{eqnarray} \label{eqn:mobility2}
\mu_{ph}= -\frac{e}{\hbar} \sum_j \langle \tau_{P j} \, v_{x j} \, \frac{\partial f\left(k_x\right)}{\partial k_x} \, \frac{1}{f\left(k_x\right)} \rangle \,
\end{eqnarray} 
where $v_{xj}=(1/\hbar) \,\, dE_j/dk_x $ is the electron velocity 
in the longitudinal direction $x$ for the $j$-th electron subband and 
$\tau_{Pj}$ is the corresponding momentum relaxation time 
for electron-phonon scattering. 
In Eq.~(\ref{eqn:mobility2}), $\langle \cdot \rangle$ denotes the expectation 
value averaged on the Fermi factor $f$ as:
\begin{eqnarray} \label{eqn:average}
\langle g \rangle= \frac{2}{n_{1D}} \int_{-\infty}^{+\infty} dk_x \, \frac{1}{2 \pi}\, g\left(k_x \right)\, f\left(k_x \right) \,
\end{eqnarray} 
where $n_{1D}$ is the one dimensional (1D) carrier density. 
In order to compute Eq.~(\ref{eqn:mobility2}), 
the following electron dispersion curve has been exploited for the $j$-th 
subband~\cite{Michetti}:
\begin{eqnarray} \label{eqn:dispersion}
E_j\left(k_x\right)= \sqrt{E_{Cj0}^2 + E_{Cj0} \frac{\hbar^2 k_x^2}{m_j}} \,+\, E_{Cj} - E_{Cj0} \, ,
\end{eqnarray} 
where $E_{Cj}= E_{Cj0}-q\Phi_C$ is the cut-off energy of the $j$-th subband 
when the electrostatic channel potential $\Phi_C$ is different from zero 
($E_{Cj}=E_{Cj0}$ for $\Phi_C=$~0~V). 
According to Ref.~\cite{Michetti}, the effective 
electron mass $m_j$ on the $j$-th subband reads:
\begin{eqnarray} \label{eqn:mass}
m_j=-\frac{2}{3}\frac{\hbar^2 E_{Cj0}}{a^2 t^2 A_j} \, , 
\end{eqnarray}
where $t$ is the graphene hopping parameter (-2.7 eV) 
and $A_j=\mbox{cos }(\pi j/(l+1))$, where $l$ is the number of dimer 
lines of the GNR. For the first conduction 
subband $E_{Cj0}=\,E_g/2$, where $E_g$ is the energy gap and $j$ 
(which runs from 1 to $l$) is the index for which $A_j$ is closest to -1/2. 

The corresponding Density of States (DOS), accounting for energy relaxation 
at outermost layers of the GNR~\cite{Michetti}, reads:
\begin{eqnarray} \label{eqn:dos}
\rho_{1Dj}(E)\!=\!\frac{2}{\pi \hbar}\sqrt{\frac{m_j \left(E+E_{Cj0}-E_{Cj} \right)^2 }{\left|E_{Cj0}\!\left(E\!-\!E_{Cj}\right)\left(E\!+\!2 E_{Cj0}\!-\!E_{Cj}\right)\right|}} \, .
\end{eqnarray} 
By means of Eqs.~(\ref{eqn:dispersion}) and~(\ref{eqn:dos}), 
the phonon-limited mobility of a 1D conductor (Eq.~(\ref{eqn:mobility2})) 
can be expressed as a sum 
over all contributing subbands $j$~\cite{Kotlyar}: 
\begin{eqnarray} \label{eqn:Kubo}
\mu_{ph} \!\!\!&=&\!\!\!\frac{2q}{\pi \hbar \, n_{2D}W k_B T}\sum_{j} \! \!\int_{E_{Cj}}^{+\infty}\!\!\!\!\!\!\!\!\!\! dE \,
\tau_{Pj}(E) \frac{f(E)\left[1\!\!-\!\!f(E)\right]}{E-E_{Cj}+E_{Cj0}}\cdot \, \nonumber  \\
&& \left(\frac{E_{Cj0}}{\,m_j}\left[\left(E-E_{Cj}+E_{Cj0}\right)^2-E_{Cj0}^2\right]\right)^{1/2} \, ,
\end{eqnarray}
where $n_{2D}=n_{1D}/W$ is the total 2D electron density, $W$ the GNR 
width and $T$ is the temperature. 

For what concerns longitudinal phonons, scattering rates are 
evaluated as in Ref.~\cite{Fang}. According to Ref.~\cite{Fang}, 
only intrasubband scattering has been considered. 
In particular, the longitudinal optical (LO) phonon scattering 
rate reads as 
\begin{eqnarray} \label{eqn:rateOPT}
1/\tau_{OP}(E)\!\!\!\!&=&\!\!\!\!\frac{n^{\mp}\pi D_{OP}^2}{4 \rho W \omega_{LO}}\rho_{1Dj}(E\pm \hbar \omega_{LO}) \nonumber \\ 
&&\left(1+\mbox{cos } \theta_{\bf k, k'}\right)\frac{1-f(E \pm \hbar \omega_{LO})}{1-f(E)} \, ,
\end{eqnarray} 
where  $n^-=1/[exp(\hbar \omega_{LO}/k_B T)-1]$ is the Bose-Einstein 
occupation factor and $n^+=n^-+1$, 
$\hbar \omega_{LO}$ is the optical phonon energy, 
$D_{OP}$ is the optical deformation potential and 
$\rho= \, 7.6 \times 10^{-8}$ g/cm$^2$ is the 2D density of graphene. 
The factor $\left(1+\mbox{cos } \theta_{\bf k, k'}\right)$ arises from 
the spinor nature of the graphene eigenfunctions and $\theta_{\bf k,k'}= 
\theta_j -\theta_{j'}$, where $\theta_j=\mbox{arctg }(k_x/k_{yj})$. Here 
$k_x$ ($k_x'$) indicates the initial (final) longitudinal electron 
wavevector referred to the Dirac point, 
whereas $k_{yj}=2 \pi j/\left[\left(l+1\right)a\right]$ and $k_{yj'}$ 
(which is equal to $k_{yj}$ for intrasubband 
scattering) are the quantized initial and final 
transverse wavevectors, respectively, where $a$ is the graphene lattice 
constant, $l$ is the number of dimer lines and $j=1,...,l$. 
The intravalley longitudinal acoustic (LA) phonon scattering rate 
can be expressed as 
\begin{equation} \label{eqn:rateAC}
1/\tau_{ACj}(E)\!=\!\frac{n_{ph}\pi D_{AC}^2 q_{x}}{4 \rho W v_S} 
\rho_{1Dj}(E)\left(\!1\!+\!\mbox{cos }\! \theta_{\bf k, k'}\!\right) \, , \nonumber
\end{equation}
where $n_{ph}=n^+ + n^-$, $D_{AC}$ is the 
deformation potential for acoustic phonons, 
$v_S=\, 2 \times 10^4$~m/s is the sound velocity in graphene and 
$|q_x|=2|k_{x}|$ is the module of the phonon wavevector 
under the backscattering condition. 

For both acoustic and optical phonons, 
we have considered the four lowest subbands. 
The electron momentum relaxation time $\tau_{Pj}$ is computed by adding the 
relaxation rate due to electron scattering with acoustic and optical 
phonons~\cite{Kotlyar}. 
As a final remark, the effective mobility including all type of 
scattering sources has been extracted 
by means of Mathiessen's rule 
$1/\mu_{tot} = 1/\mu_{LER}+1/\mu_{d}+1/\mu_{IMP}+1/\mu_{ph}$, 
where $\mu_d$ and $\mu_{IMP}$ are the defect and impurity limited mobilities, 
respectively. 
We have verified the validity of Mathiessen's rule  
considering samples with more sources of non-idealities (i.e. LER, ionized 
impurities and defects) at the same time. Then we have compared the computed mobility 
with that obtained by adding single contributions with Mathiessen's rule, observing a 
relative error smaller than 3\%, which lies within the statistical error.
 
\section{Results and Discussions} 

\begin{figure} [tbp] 
\begin{center} 
\includegraphics[width=7.0cm]{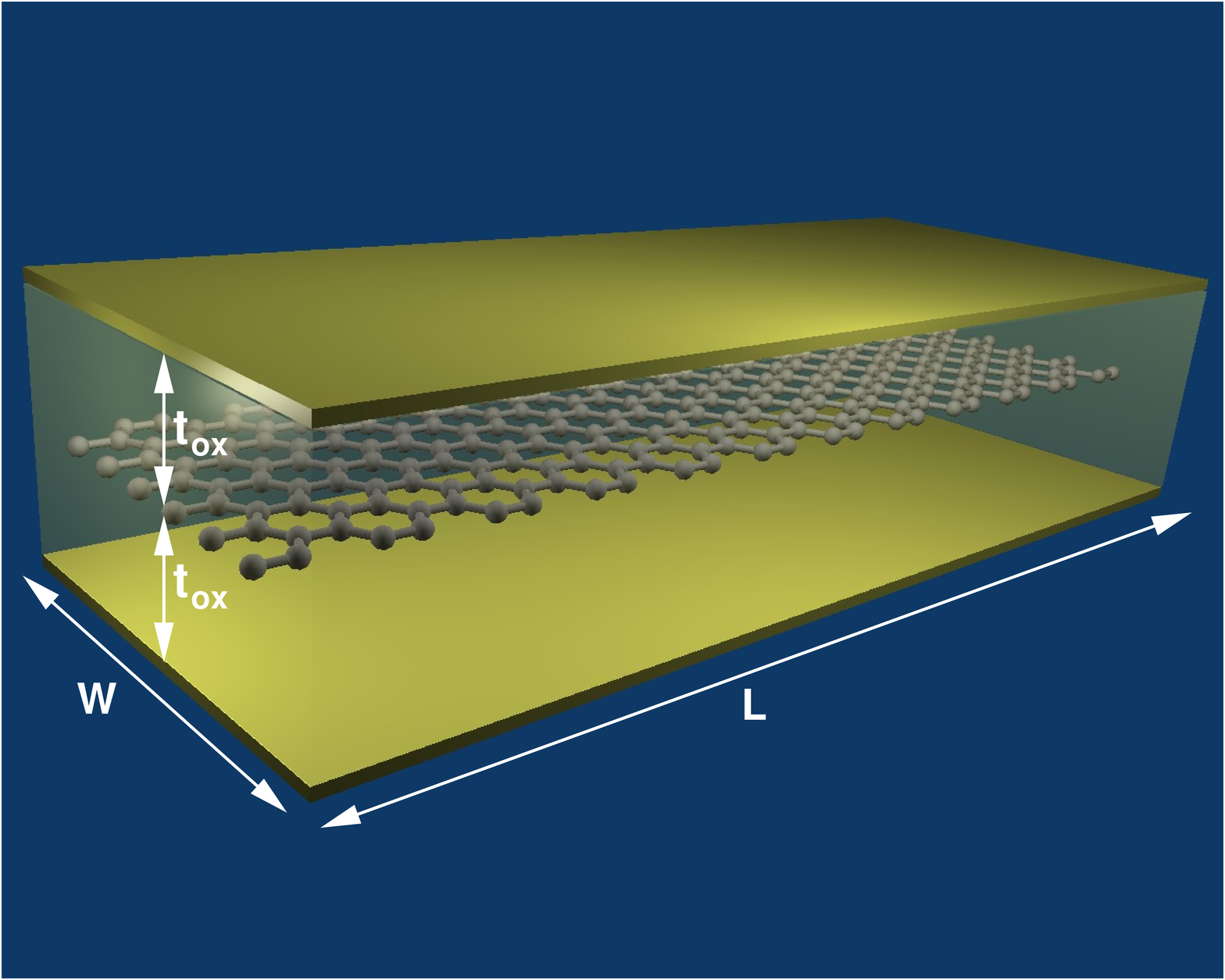}
\end{center}
\vspace{-0.2cm}
\caption{3D structure of the simulated GNR segment.} 
\label{fig:device}
\end{figure}

The simulated segment is a 
double-gate GNR, embedded in SiO$_2$ with an oxide thickness $t_{ox}$ 
of 2~nm, 10~nm-long (Fig.~\ref{fig:device}). 
The segment length has been chosen to satisfy the assumption of 
loss of phase coherence at the segment ends. Indeed, 
according to recent experiments~\cite{CasiraghiNL}, 
the phase coherence length is close to 11~nm in graphene. 
From a computational point of view, different widths $W$ have 
been considered, ranging from 1 to 10~nm: 
1.12 nm, 2.62 nm, 4.86 nm and 10.10 nm. 
All simulations have been performed at room temperature $T =300$~K. 

\subsection{Line-edge roughness limited mobility} 

\begin{figure} [tbp] 
\begin{center} 
\includegraphics[width=8.6cm]{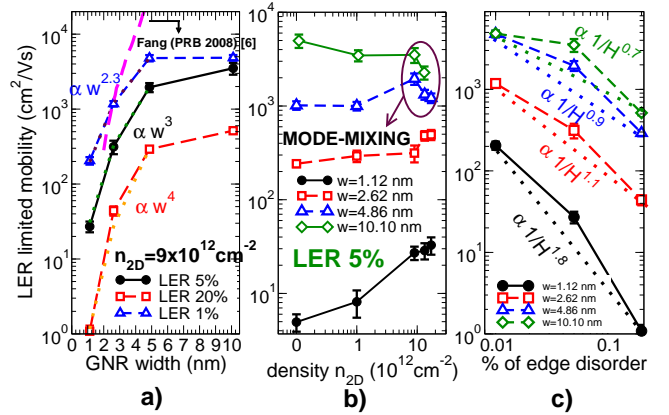}
\end{center}
\vspace{-0.6cm}
\caption{a) LER-limited mobility as a function of $W$ for $n_{2D}=0.9\times 10^{13}$~cm$^{-2}$ and for different $H$. Data extracted from Ref.~\cite{Fang} 
are also reported. b) LER-limited mobility as a 
function of $n_{2D}$ for $H=5\%$. c) LER-limited mobility as a function of edge disorder concentration $H$ for
$n_{2D}=0.9 \times 10^{13}$~cm$^{-2}$ and for different GNR width $W$.}
\label{fig:LER}
\end{figure}

LER-limited mobility as a function 
of $W$ for different edge-defect concentrations $H$ is 
shown in Fig.~\ref{fig:LER}a in the above-threshold regime, for a 2D carrier 
density $n_{2D}$ of $9\times 10^{12}$ cm$^{-2}$. 
As in all figures in the paper, the error bars represent the estimated root 
mean squared error $\sigma_{\mu}$ of the average of the statistical sample 
(\ref{eqn:error1}). 

As predicted by the analytical model in Ref.~\cite{Fang}, 
$\mu_{LER}$ scales as $W^4$. 
Such behavior holds for large $H$ ($\approx$20\%) and narrow GNRs 
($W<$ 5~nm), when scattering from edge defects is expected to be heavier, 
while, for wider GNRs and for smaller $H$, such a law is not obeyed. 
In particular, for GNR width larger than 5~nm, 
$\mu_{LER}$ tends to saturate, since the increasing 
number of subbands contributing to transport counterbalance the 
number of final states available for scattering, enhancing scattering rates. 
As shown in Fig.~\ref{fig:LER}b, in narrower GNRs, 
the higher the electron density, the larger the effective 
mobility, because of stronger screening. 
$\mu_{LER}$ decreases for high $n_{2D}$ and wider GNRs, due to mode-mixing, as 
already observed in Silicon Nanowire FETs~\cite{Pala}. 
Indeed, for wider GNRs biased in the inversion regime, more transverse 
modes are able to propagate in the channel due to the reduced energy 
separation between different subbands. 
This leads edge defects to 
become a source of intermode scattering, thus reducing $\mu_{LER}$. 

Fig.~\ref{fig:LER}c shows $\mu_{LER}$ as a function of $H$, where 
$\mu_{LER} \propto 1/H$ for wide GNRs, consistent with 
the Drude model, and also observed in graphene 
in the presence of defects~\cite{Chen1}. 
However, as soon as $W$ decreases, quantum localization becomes 
relevant~\cite{Evaldsson}, and the Anderson insulator-like 
behavior~\cite{Querlioz} is recovered ($\mu_{LER} \propto 1/L^2$), 
in agreement with analytical predictions~\cite{Fang}. 

\subsection{Defect-limited mobility} 

\begin{figure} [tbp] 
\begin{center} 
\includegraphics[width=8.5cm]{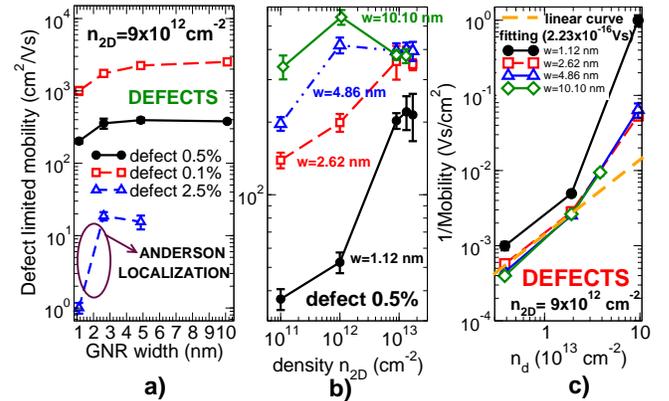}
\end{center}
\vspace{-0.6cm}
\caption{a) Defect-limited mobility as a function of $W$ for
$n_{2D}=\, 9 \times 10^{12}$~cm$^{-2}$ and for different 
defect fraction $n_{d}$. b) Mobility as a function of $n_{2D}$ for a defect 
fraction $n_d=$ 0.5\%. c) Inverse of the mobility as a function of $n_d$ for 
$n_{2D}=\, 9 \times 10^{12}$~cm$^{-2}$ and for different GNR widths $W$. } 
\label{fig:defects}
\end{figure}

Defect-limited mobility is plotted 
in Fig.~\ref{fig:defects}a as a function of $W$ for different defect 
concentrations. 
Even in this case, localization affects mobility in narrower ribbons, 
especially for higher $n_d$ (2.5\%). 
\begin{figure} [tbp] 
\begin{center} 
\includegraphics[width=8.5cm]{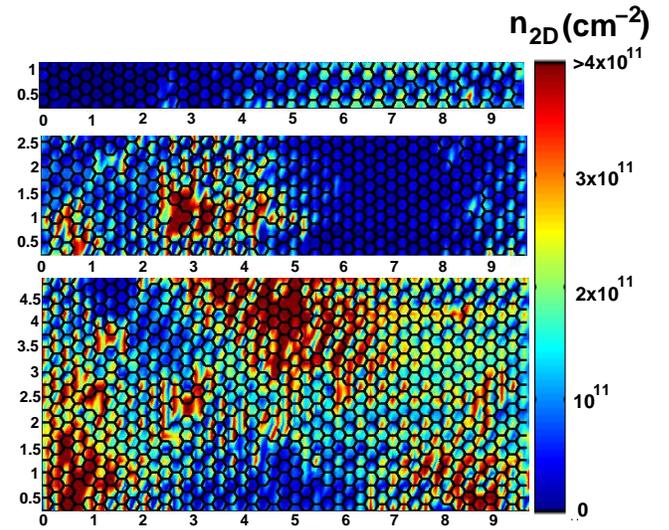}
\end{center}
\vspace{-0.2cm}
\caption{Carrier density $n_{2D}$ in the GNR channel for different GNR width: 
$W=$~1.12~nm, 2.62~nm and 4.86~nm (from top to bottom).} 
\label{fig:Anderson}
\end{figure} 

For a fixed defect density, mobility slightly increases with electron density, 
due to the larger screening (Fig.~\ref{fig:defects}b) and, for larger GNRs 
biased in the inversion regime, it saturates with increasing $W$, for 
the same reason discussed above for LER scattering. 
In Fig.~\ref{fig:defects}c, $\mu_{d}$ is plotted as a function of $n_d$. 
The wider the ribbons the closer mobility follows the simple Drude 
model ($\mu_{d} \propto 1/n_d$), as expected for strong 
disorder and uncorrelated scatterers in 2D graphene sheets~\cite{Stauber}. 
For $W=$~10.10~nm atomistic 
simulations are in agreement with experimental results: 
a linear curve fitting ($\mu= \,C/n_d$) leads to a proportionality factor 
of $2.23\times 10^{-16}$~Vs, similar to those extracted in the case of Ne$^+$ 
and He$^+$ irradiated graphene samples ($7.9\times 10^{-16}$~Vs 
and $9.3\times 10^{-16}$~Vs, respectively)~\cite{Chen1}.

In Fig.~\ref{fig:Anderson}, the GNR carrier density for widths ranging 
from 1.12~nm to 4.86~nm is shown. 
As can be seen, in Fig.~\ref{fig:Anderson}, Anderson localization strongly 
degrades electron mobility~\cite{Evaldsson}, creating percolating paths 
in wider GNRs and blocking conduction in the narrower ones. 

\subsection{Ionized impurities limited mobility} 

\begin{figure} [t] 
\begin{center} 
\includegraphics[width=8.6cm]{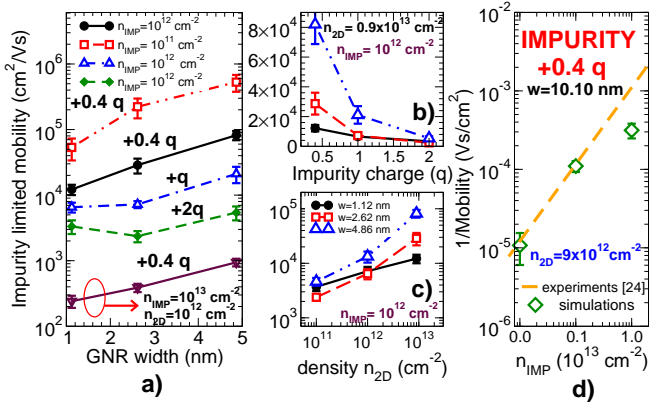}
\end{center}
\vspace{-0.5cm}
\caption{a) Impurity-limited mobility as a function of $W$ for different 
impurity concentrations $n_{IMP}$ and impurity charges. 
b) Mobility as a function of the impurity charge for 
$n_{IMP}=\, 10^{12}$~cm$^{-2}$ and for different $W$. In a) and b) 
$n_{2D}=\,0.9\times 10^{13}$~cm$^{-2}$, except otherwise specified. 
c) Impurity-limited mobility as a function of $n_{2D}$ for different 
$W$ ($n_{IMP}=10^{12}$~cm$^{-2}$ and the impurity charge is $+0.4q$). 
d) Inverse mobility as a function of $n_{IMP}$ for $W=$~10.10~nm. 
The carrier density is $n_{2D}=\,0.9\times 10^{13}$~cm$^{-2}$. The 
experimental slope $2 \times 10^{-16}$~Vs extracted in 
Ref.~\cite{Chen2} is also reported.} 
\label{fig:impurity}
\end{figure}

Impurity-limited mobility $\mu_{\rm IMP}$, as a function
of $W$, is shown in Fig.~\ref{fig:impurity}a 
for $n_{2D}=9\times 10^{12}$~cm$^{-2}$, and for different impurity charge 
concentrations. As can be noted, even a high impurity concentration of 
10$^{12}$ cm$^{-2}$ yields large mobility for 0.4$q$ impurity charge. 
However, no indications are present in 
literature regarding the amount of unintentional doping 
charge~\cite{Casiraghi,Chen2}. Therefore, 
in order to check also the effect of impurity ionization on the electron 
transport, statistical simulations have been performed by increasing the 
impurity charge up to +2$q$. 
Mobility as a function of the impurity charge is plotted
 in Fig.~\ref{fig:impurity}b for different $W$ and for 
$n_{2D}=9\times 10^{12}$~cm$^{-2}$. 
In this case smaller values of $\mu$ 
(1700 cm$^2$/Vs) are obtained for very narrow GNRs, due to the 
strongly nonlinear impact on screening in the channel.
Even in this case localization strongly degrades mobility for narrower 
ribbons. 
\begin{figure} [tbp] 
\begin{center} 
\includegraphics[width=8.6cm]{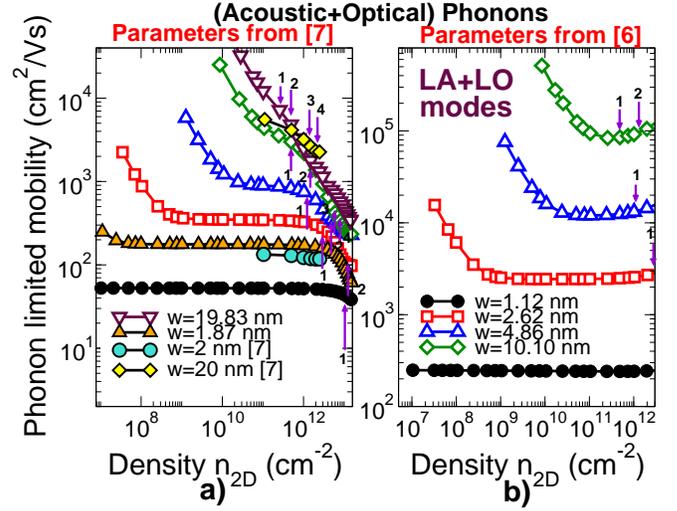}
\end{center}
\vspace{-0.5cm}
\caption{a) Mobility limited by phonons 
(zone-boundary ZO+acoustic) scattering as a function of 
$n_{2D}$ for different $W$, computed by means of 
parameters from~\cite{Bresciani}. 
Data from Ref.~\cite{Bresciani} are also reported. 
b) Same plot of a) exploiting 
parameters from~\cite{Fang}, corresponding to zone-boundary 
LO mode. 
In both plots the threshold densities at which the different subbands are 
activated are sketched.}
\label{fig:phonons}
\end{figure} 
\begin{figure} [tbp] 
\begin{center} 
\includegraphics[width=8.6cm]{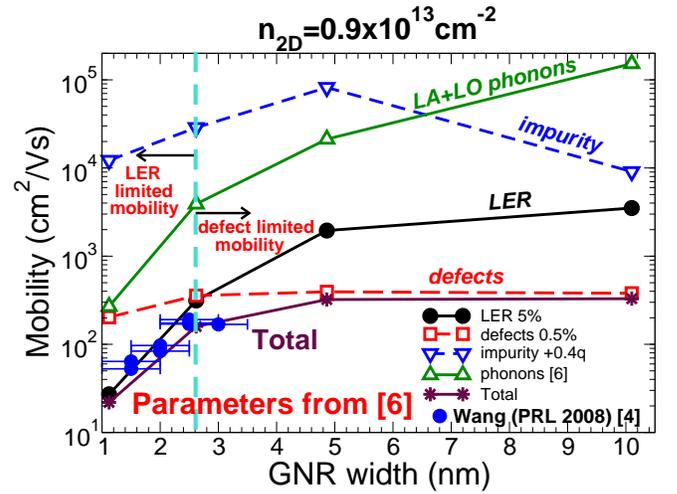}
\end{center}
\vspace{-0.5cm}
\caption{Mobility limited by phonons, LER, defect and impurity 
scattering in the inversion regime for a LER concentration $H$=5\%, 
$n_d=$ 0.5\%. The parameters for the scattering rates have been taken 
from~\cite{Fang}~. The experimental mobility from~\cite{Wang} is 
also reported. $n_{2D}=0.9 \times 10^{13}$~cm$^{-2}$, 
$n_{IMP}=\,10^{12}$ cm$^{-2}$.}
\label{fig:mutot}
\end{figure} 

To further test the importance of unintentional doping in limiting 
mobility, we have considered excess charge densities up to 
10$^{13}$~cm$^2$, which have been encountered in experiments~\cite{Casiraghi}. 
As shown in Fig.~\ref{fig:impurity}a, in this case mobility decreases down 
to 10$^2$~cm$^2$/Vs for narrower GNRs. In Fig.~\ref{fig:impurity}c 
impurity-limited mobility is plotted 
as a function of $n_{2D}$ for $n_{IMP}=10^{12}$~cm$^{-2}$ and impurity charge 
$+0.4q$ and for different $W$. 
According to~\cite{Hwang,Bolotin}, $\mu_{\rm IMP}$ in graphene does not 
depend on the electron density. The behavior is different in GNRs because
up to an electron density of  
10$^{12}$~cm$^{-2}$, only the ground state is occupied, so that the Size 
Quantum Limit approximation is verified~\cite{Lee,Fang}. 
Since the scattering rate $1/\tau \propto \epsilon^{-2} $~\cite{Hwang} and 
the static dielectric function $\epsilon$ increases with 
$n_{2D}$~\cite{Lee}, the screening becomes stronger 
with increasing $n_{2D}$. As a consequence, $\mu_{IMP} \propto \tau$ in GNRs
has the increasing monotonic behavior shown in Fig.~\ref{fig:impurity}c. 
In Fig.~\ref{fig:impurity}d, we compare experimental results 
available in literature~\cite{Chen2} for graphene, 
showing the inverse of the impurity-limited mobility as a function of 
$n_{IMP}$ for $W=$~10.10~nm and by considering an impurity charge of +0.4$q$: 
as expected for uncorrelated 
scatterers, $\mu_{\rm IMP} \propto 1/n_{IMP}$ and, 
as can be seen, experiments and simulations show quite good agreement. 

\subsection{Acoustic and optical phonon-limited mobility} 
Our study has also 
been directed towards the investigation of the impact of phonon scattering, 
through the Kubo-Greenwood formalism~\cite{Kubo,Greenwood}. 
A wide range of phonon parameter values 
is currently present in the literature~\cite{Chen,Borysenko,Fang,Bolotin} 
(i.e. acoustic ($D_{AC}$) and optical ($D_{OP}$) 
deformation potentials, as well as optical phonon energy $\hbar \omega_{LO}$). 
We observe that the most widely used 
phonon parameters are those adopted in Refs.~\cite{Fang,Chen,Obradovic}, 
i.e. $D_{AC}=$~16 eV, $\hbar \omega_{LO}=$~160 meV and 
$D_{OP}=1.4\times 10^9$~eV/cm, where $\hbar \omega_{LO}$ is the 
zone-boundary LO phonon energy. 
Such values have been tested towards those 
provided in Refs.~\cite{Bresciani,Shishir,Lazzeri}, showing good agreement as 
far as mobility is concerned.  

To prove the validity of our approach, we have first compared our 
results with those obtained by means of an accurate 2D Monte Carlo 
simulator~\cite{Bresciani}. For a fair comparison, the same phonon 
parameters and the same scattering rates as in~\cite{Bresciani} have 
been used. 
As can be seen in Fig.~\ref{fig:phonons}a, results are in good agreement, 
especially for wider GNRs. 
However, such parameters correspond to the out-of plane mode ZO which, 
according to symmetry-based considerations~\cite{Manes}, 
density functional study~\cite{Malola} and 
Raman spectroscopy~\cite{Ferrari}, is much weaker 
than in-plane vibrations. 
 
Therefore, in the following, we adopt the parameters discussed above for 
LA and LO phonons and the scattering rates described in 
Eqs.~(\ref{eqn:rateOPT}) and~(\ref{eqn:rateAC}).  
In Fig.~\ref{fig:phonons}b acoustic and optical phonon limited mobility 
is shown as a function of $n_{2D}$. As expected, emission scattering 
rates are found to be larger than absorption scattering rates, 
due to their higher Bose-Einstein occupation numbers. In addition, 
as also observed in graphene~\cite{Akturk}, 
we have verified that the contribution of optical phonons 
is negligible also in GNRs, and $\mu_{ph}$ is dominated by (intravalley) 
acoustic phonon scattering~\cite{Fang,Bresciani} 
(Fig.~\ref{fig:phonons}b). Note also that, unlike in grafene where 
$\mu_{ph} \! \propto \! 1/n_{2D}$~\cite{Perebeinos}, 
in GNRs the transverse confinement 
leads to a non-monotonic $n_{2D}$-dependence as in CNTs~\cite{PerebeinosNL}. 
As can be seen, $\mu_{ph}$ slightly increases due to the reduced number 
of available states for scattering. 

We observe that several recent studies~\cite{Perebeinos,Fratini,Konar} 
have demonstrated that 
surface phonons of the substrate represent a severe source of scattering, 
which strongly limits transport in graphene. However, we expect this 
effect to be much larger in high-$k$ dielectrics like HfO$_2$, rather than in 
SiO$_2$, which is the insulator considered in this work. 
This issue will be the topic of a more comprehensive work on 
electron-phonon scattering in GNRs, which is beyond the scope of the present paper.  \\ \\
Finally, we compare 
the total mobility with experiments from Wang 
et al~\cite{Wang}~(Fig.~\ref{fig:mutot}). In particular, we show the 
mobility limited by different scattering mechanisms as well as 
the total mobility computed by means of Mathiessen's rule. 
As can be seen, when using parameters in Ref.~\cite{Fang}, LER is the 
most limiting mechanism ($H=$ 5\%) for very narrower GNRs, while for wider 
GNRs defect scattering is predominant, if a $n_d=$ 0.5\% is considered. 
As an additional remark, we have checked that the same conclusion holds even if we consider
much lower deformation potentials for phonons, that decrease the impact of phonon scattering, such as those
those provided in Ref. \cite{Borysenko}.

\section{Conclusion} 
We have defined a simulation methodology 
based on atomistic simulations on statistically significant ensembles of 
GNR segments to understand the functional dependence of GNR mobility upon 
different factors, and to quantitatively assess the importance of 
different scattering mechanisms. 

We used such methodology to investigate mobility in GNRs of width ranging 
from 1 to 10 nm.
First, we find that, unlike in 2D graphene, electron-impurity 
scattering in GNRs is far too weak to affect low-field mobility. 
In addition, using well established parameters for electron-phonon coupling,
we find that phonon scattering is hardly the limiting factor of GNR mobility.
For narrower GNRs, line-edge roughness is the main scattering mechanism.
This result is consistent with the findings in~\cite{Yang}, 
where wider nanoribbons 
with very rough edges are characterized.
Finally, for a fixed defect density or LER, mobility tends to decrease with 
the GNR width for narrower devices, suggesting the occurrence of localization effects. 

\section{Acknowledgment}
Authors thank M. Lemme, P. Palestri, 
P. Michetti and T. Fang for useful 
discussions, and CINECA Super-Computing Center, Bologna, and 
www.nanohub.org for the provided computational resources.
This work was supported in part by the EC 7FP 
through the Network of Excellence NANOSIL (Contract 216171), 
and the GRAND project (Contract 215752), by MIUR through the PRIN GRANFET project (Prot. 200852CLJ9), 
by the ESF EUROCORES Programme FoNE, through funding from the CNR (awarded to IEEIIT-PISA) and the EC 6FP, 
under Project Dewint (Contract ERASCT-2003-980409). 


\begin{thebibliography}{10}

\bibitem{Chen}
J.-H. Chen, C.~Jang, S.~Xiao, M.~Ishigami, and M.~S. Fuhrer.
\newblock Intrinsic and extrinsic performance limits of graphene devices on
  sio$_2$.
\newblock {\em Nature Nanotechnology}, vol. 3, pp. 206--209, 2008.

\bibitem{Li}
X.~Li, X.~Wang, L.~Zhang, S.~Lee, and H.~Dai.
\newblock Chemically derived, ultrasmooth graphene nanoribbon semiconductors.
\newblock {\em Science}, vol. 319, pp. 1229--1231, 2008.

\bibitem{Appenzeller}
Z.~Chen and J.~Appenzeller.
\newblock Mobility {E}xtraction and {Q}uantum {C}apacitance {I}mpact in {H}igh
  {P}erformance {G}raphene {F}ield-effect {T}ransistor {D}evices.
\newblock {\em IEDM Tech. Digest}, pages 509--512, 2008.

\bibitem{Wang}
X.~Wang, Y.~Ouyang, X.~Li, H.~Wang, J.~Guo, and H.~Dai.
\newblock Room-temperature all-semiconducting sub-10 nm graphene nanoribbon
  field-effect transistors.
\newblock {\em Phys. Rev. Lett.}, vol. 100, pp. 206803, 2008.

\bibitem{Yang}
Y.~Yang and R.~Murali.
\newblock Impact of {S}ize {E}ffect on {G}raphene {N}anoribbon {T}ransport.
\newblock {\em IEEE Elec. Dev. Lett.}, vol. 31, pp. 237--239, 2010.

\bibitem{Fang}
T.~Fang, A.~Konar, H.~Xing, and D.~Jena.
\newblock Mobility in semiconducting graphene nanoribbons: Phonon, impurity,
  and edge roughness scattering.
\newblock {\em Phys. Rev. B}, vol. 78, pp. 205403, 2008.

\bibitem{Bresciani}
M.~Bresciani, P.~Palestri, and D.~Esseni.
\newblock Simple end efficient modeling of the {E}-k relationship and low-field
  mobility in {G}raphene {N}ano-{R}ibbons.
\newblock {\em Solid-State Electronics}, vol. 54, pp. 1015--1021, 2010.

\bibitem{Areshkin}
D.~A. Areshkin, D.~Gunlycke, and C.~T. White.
\newblock Ballistic {T}ransport in {G}raphene {N}anostrips in the {P}resence of
  {D}isorder: {I}mportance of {E}dge {E}ffects.
\newblock {\em Nano Lett.}, vol. 7, pp. 204--210, 2007.

\bibitem{Querlioz}
D.~Querlioz, Y.~Apertet, A.~Valentin, K.~Huet, A.~Bournel,
  S.~Galdin-Retailleau, and P.~Dollfus.
\newblock Suppression of the orientation effects on bandgap in graphene
  nanoribbons in the presence of edge disorder.
\newblock {\em Appl. Phys. Lett.}, vol. 92, pp. 042108, 2008.

\bibitem{BettiIEDM09}
A.~Betti, G.~Fiori, G.~Iannaccone, and Y.~Mao.
\newblock Physical insights on graphene nanoribbon mobility through atomistic
  simulations.
\newblock {\em IEDM Tech. Digest}, pages 897--900, 2009.

\bibitem{IannaPala}
G.~Iannaccone and M.~Pala.
\newblock An extended concept of mobility for non diffusive conductors,
\newblock {\em unpublished}.

\bibitem{ViDES}
Code and {D}ocumentation can be found at the url:
  http://www.nanohub.org/tools/vides.

\bibitem{Deretzis}
I.~Deretsiz, G.~Forte, A.~Grassi, A.~La Magna, G.~Piccitto, and R.~Pucci.
\newblock A multiscale study of electronic structure and quantum transport in
  ${C}_{6n^2}{H}_{6n}$-based graphene quantum dots.
\newblock {\em J. Phys.: Condens. Matter}, vol. 22, pp. 095504, 2010.

\bibitem{Rytkonen}
K.~Rytk\"onen, J.~Akola, and M.~Manninen.
\newblock Density functional study of alkali metal atoms and monolayers on
  graphite (0001).
\newblock {\em Phys. Rev. B}, vol. 75, pp. 075401, 2007.

\bibitem{Michetti}
P.~Michetti and G.~Iannaccone.
\newblock Analytical {M}odel of {O}ne-{D}imensional {C}arbon-{B}ased
  {S}chottky-{B}arrier {T}ransistors.
\newblock {\em IEEE Trans. Electron Devices}, vol. 57, pp. 1616--1625, 2010.

\bibitem{Fischetti}
M.~Fischetti.
\newblock {ECE}609 {P}hysics of {S}emiconductor {D}evices ({S}pring 2010).
\newblock {\em http://www.ecs.umass.edu/ece/ece609}, Part 1, pp. 80, 2010.

\bibitem{Kotlyar}
R.~Kotlyar, B.~Obradovic, P.~Matagne, M.~Stettler, and M.~D. Giles.
\newblock Assessment of room-temperature phonon-limited mobility in gated
  silicon nanowires.
\newblock {\em Appl. Phys. Lett.}, vol. 84, pp. 5270, 2004.

\bibitem{CasiraghiNL}
C.~Casiraghi, A.~Hartschuh, H.~Qian, S.~Piscanec, C.~Georgi, A.~Fasoli, K.~S.
  Novoselov, D.~M. Basko, and A.~C. Ferrari.
\newblock {R}aman {S}pectroscopy of {G}raphene {E}dges.
\newblock {\em Nano Lett.}, vol. 9, pp. 1433--1441, 2009.

\bibitem{Pala}
S.~Poli, M.~G. Pala, T.~Poiroux, S.~Deleonibus, and G.~Baccarani.
\newblock {S}ize {D}ependence of {S}urface-{R}oughness-{L}imited {M}obility in
  {S}ilicon-{N}anowire {F}{E}{T}s.
\newblock {\em IEEE Trans. Electron Devices}, vol. 55, pp. 2968--2976, 2008.

\bibitem{Chen1}
J.-H. Chen, W.~G. Cullen, C.~Jang, M.~S. Fuhrer, and E.~D. Williams.
\newblock Defect {S}cattering in {G}raphene.
\newblock {\em Phys. Rev. Lett.}, vol. 102, pp. 236805, 2009.

\bibitem{Evaldsson}
M.~Evaldsson, I.~V. Zozoulenko, H.~XU, and T.~Heinzel.
\newblock Edge-disorder-induced anderson localization and conduction gap in
  graphene nanoribbons.
\newblock {\em Phys. Rev. B}, vol. 78, pp. 161407, 2008.

\bibitem{Stauber}
T.~Stauber, N.~M.~R. Peres, and F.~Guinea.
\newblock Electronic transport in graphene: {A} semiclassical approach
  including midgap states.
\newblock {\em Phys. Rev. B}, vol. 76, pp. 205423, 2007.

\bibitem{Casiraghi}
C.~Casiraghi, S.~Pisana, K.~S. Novoselov, A.~K. Geim, and A.~C. Ferrari.
\newblock Raman fingerprint of charged impurities in graphene.
\newblock {\em Appl. Phys. Lett.}, vol. 91, pp. 233108, 2007.

\bibitem{Chen2}
J.-H. Chen, C.~Jang, S.~Adam, M.~S. Fuhrer, E.~D. Williams, and M.~Ishigami.
\newblock Charged-impurity scattering in graphene.
\newblock {\em Nature Physics}, vol. 4, pp. 377--381, 2008.

\bibitem{Hwang}
E.~H. Hwang, S.~Adam, and S.~Das Sarma.
\newblock {C}arrier {T}ransport in {T}wo-{D}imensional {G}raphene {L}ayers.
\newblock {\em Phys. Rev. Lett.}, vol. 98, pp. 186806, 2007.

\bibitem{Bolotin}
K.~I. {B}olotin~et al.
\newblock {T}emperature-{D}ependent {T}ransport in {S}uspended {G}raphene.
\newblock {\em Phys. Rev. Lett.}, vol. 101, pp. 096802, 2008.

\bibitem{Lee}
J.~Lee and H.~Spector.
\newblock Dielectric response function for a quasi-one-dimensional
  semiconducting system.
\newblock {\em J. Appl. Phys.}, vol. 57, pp. 366--372, 1985.

\bibitem{Kubo}
R.~Kubo.
\newblock Statistical-{M}echanical {T}heory of {I}rreversible {P}rocesses.
\newblock {\em J. Phys. Soc. Jpn.}, vol. 12, pp. 570, 1957.

\bibitem{Greenwood}
D.~A. Greenwood.
\newblock The {B}oltzmann {E}quation in the {T}heory of {E}lectrical
  {C}onduction in {M}etals.
\newblock {\em Proc. Phys. Soc. London}, vol. 71, pp. 585, 1958.

\bibitem{Borysenko}
K.~M. Borysenko, J.~T. Mullen, E.~A. Barry, S.~Paul, Y.~G. Semenov, J.~M.
  Zavada, M.~Buongiorno Nardelli, and K.~W. Kim.
\newblock First-principles analysis of electron-phonon interactions in
  graphene.
\newblock {\em Phys. Rev. B}, vol. 81, pp. 121412, 2010.

\bibitem{Obradovic}
B.~Obradovic, R.~Kotlyar, F.~Heinz, P.~Matagne, T.~Rakshit, M.~D. Giles, and
  M.~A. Stettler.
\newblock Analysis of graphene nanoribbons as a channel material for
  field-effect transistors.
\newblock {\em Appl. Phys. Lett.}, vol. 88, pp. 142102, 2006.

\bibitem{Shishir}
R.~S. Shishir and D.~K. Ferry.
\newblock Intrinsic mobility in graphene.
\newblock {\em J. Phys.: Condens. Matter}, vol. 21, pp. 232204, 2009.

\bibitem{Lazzeri}
M.~Lazzeri, S.~Piscanec, F.~Mauri, A.~C. Ferrari, and J.~Robertson.
\newblock Electron {T}ransport and {H}ot {P}honons in {C}arbon {N}anotubes.
\newblock {\em Phys. Rev. Lett.}, vol. 95, pp. 236802, 2005.

\bibitem{Manes}
J.~L. Manes.
\newblock Symmetry-based approach to electron-phonon interactions in graphene.
\newblock {\em Phys. Rev. B}, vol. 76, pp. 045430, 2007.

\bibitem{Malola}
S.~Malola, H.~H\"akkinen, and P.~Koskinen.
\newblock {C}omparison of {R}aman spectra and vibrational density of states
  between graphene nanoribbons with different edges.
\newblock {\em Eur. Phys. Journ. D}, vol. 52, pp. 71--74, 2008.

\bibitem{Ferrari}
A.~C. {F}errari~et al.
\newblock {R}aman {S}pectrum of {G}raphene and {G}raphene {L}ayers.
\newblock {\em Phys. Rev. Lett.}, vol. 97, pp. 187401, 2006.

\bibitem{Akturk}
A.~Akturk and N.~Goldsman.
\newblock Electron transport and full-band electron-phonon interactions in
  graphene.
\newblock {\em J. Appl. Phys.}, vol. 103, pp. 053702, 2008.

\bibitem{Perebeinos}
V.~Perebeinos and P.~Avouris.
\newblock Inelastic scattering and current saturation in graphene.
\newblock {\em Phys. Rev. B}, vol. 81, pp. 195442, 2010.

\bibitem{PerebeinosNL}
V.~Perebeinos, J.~Tersoff, and P.~Avouris.
\newblock Mobility in {S}emiconducting {C}arbon {N}anotubes at {F}inite
  {C}arrier {D}ensity.
\newblock {\em Nano Lett.}, vol. 6, pp. 205--208, 2006.

\bibitem{Fratini}
S.~Fratini and F.~Guinea.
\newblock Substrate-limited electron dynamics in graphene.
\newblock {\em Phys. Rev. B}, vol. 77, pp. 195415, 2008.

\bibitem{Konar}
A.~Konar, T.~Fang, and D.~Jena.
\newblock Effect of high-$k$ gate dielectrics on charge transport in
  graphene-based field effect transistors.
\newblock {\em Phys. Rev. B}, vol. 82, pp. 115452, 2010.

\end{thebibliography}

\end{document}